\documentclass[sigconf,nonacm]{acmart}
\AtBeginDocument{%
  }

\usepackage{algorithm}
\usepackage{algpseudocode}
\usepackage{amsmath} 
\usepackage{subcaption}
\usepackage{tabularx}
\usepackage{booktabs}
\usepackage{multirow}

\floatname{algorithm}{Prompt}

\renewcommand\footnotetextcopyrightpermission[1]{}
\AtBeginDocument{%
  \pagestyle{plain}%
}

\begin{document}

\title{Modeling Liquid-Cooled Data Centers in EnergyPlus}


\author{Yun-Dam Ko}
\authornote{Equal contribution.}
\email{yundamko@stanford.edu}
\orcid{0009-0001-1647-496X}
\affiliation{%
  \institution{Stanford University}
  \city{Stanford}
  \state{California}
  \country{USA}
}

\author{Dipashreya A. Sur}
\authornotemark[1]
\email{dipashre@stanford.edu}
\orcid{0000-0002-7418-3415}
\affiliation{%
  \institution{Stanford University}
  \city{Stanford}
  \state{California}
  \country{USA}
}

\author{Cade Napier}
\authornotemark[1]
\email{cnapier@stanford.edu}
\orcid{}
\affiliation{%
  \institution{Stanford University}
  \city{Stanford}
  \state{California}
  \country{USA}
}

\author{Carlos D. D{\'{\i}}az-Mar{\'{\i}}n}
\email{cdiazm@stanford.edu}
\orcid{}
\affiliation{%
  \institution{Stanford University}
  \city{Stanford}
  \state{California}
  \country{USA}
}

\author{Rishee K. Jain}
\email{rishee.jain@stanford.edu}
\orcid{}
\affiliation{%
  \institution{Stanford University}
  \city{Stanford}
  \state{California}
  \country{USA}
}


\renewcommand{\shortauthors}{Ko et al.}

\begin{abstract}
As AI workloads accelerate data center transition toward high-density racks with direct-to-chip (DTC) liquid cooling, existing simulation tools remain limited in evaluating whole-building and multi-year efficiency. To bridge this gap, we develop a native EnergyPlus object, \texttt{ElectricEquipment:\allowbreak ITE:\allowbreak LiquidCooled}, that models DTC liquid cooling for chilled-water and warm-water architectures as plant-loop components. It couples facility-water loops, rack-side secondary loops, coolant distribution units, and chip-junction temperatures within a single solver. Validation against an independent Modelica implementation across 21 operating conditions, including two full-year runs compared at one-minute resolution, yields loop-temperature RMSE at or below 0.021~K, well within the 0.1~K acceptance threshold. Our case study demonstrates that warm-water liquid cooling reduces annual PUE from 1.43 to 1.25, eliminates on-site water use, and cuts total facility energy by 37.9\% relative to an air-cooled baseline. By embedding liquid-cooling logic into EnergyPlus, our work provides whole-building, long-term energy modeling for high-density AI infrastructure within an open-source framework.

\end{abstract}


\maketitle

\section{Introduction}
\label{sec:introduction}
Global data center energy consumption is rising at an unprecedented rate necessitating modeling to identify sources of energy waste. Estimates indicate that data center energy use could contribute to 9.5-15.3\% of U.S. electricity consumption by 2030~\cite{LBNL2026}. In parallel, the scale of new data center campuses is also increasing as over 50\% are predicted to exceed 500 MW by 2035, and nearly one-third to exceed 1 GW~\cite{bloom2026power}. Therefore, modeling data centers holistically can identify optimal building strategies that lower their power demands and make them more energy efficient and flexible. 

As data centers continue expanding to meet growing artificial intelligence (AI) demands, new liquid cooling methodologies need to be explored and deployed. Traditional air cooling uses computer room air handler (CRAH) units to circulate and cool warm air through server rack arrangements. Liquid cooling, by contrast, circulates coolant directly through components to absorb and dissipate heat. These approaches differ significantly in their suitability for supporting AI demands. Traditional air-cooled server racks operate at 7-10 kW, while AI computing racks can reach 30-100+ kW per rack~\cite{CHEN2026100162, nlyte2024rackpower}. AI compute chips in these racks can reach heat fluxes up to $106\ \mathrm{W/cm^2}$~\cite{ren2026experimental}, making air cooled systems no longer viable given the high power density~\cite{coolinglimit2022}.

Tools capable of modeling data-center liquid cooling at the building level already exist. The Modelica Buildings library~\cite{WetterBuildingsLibrary} supports high-fidelity simulation of liquid-cooling equipment alongside its own thermal-zone and envelope models, and can additionally be coupled to the EnergyPlus engine through Spawn-of-EnergyPlus~\cite{Spawn03032024}. Its use at the facility scale nevertheless remains limited in practice. Modelica has a steep learning curve despite decades of software development and focuses more on equipment-level components, limiting its usage for multi-year facility scale runs~\cite{qiu2024review}. Therefore, it remains challenging to assess how liquid cooling interacts with the data center design. Extending the modeling capabilities at the facility level, using tools such as EnergyPlus, could allow for identification of how liquid cooling interacts with heat recovery systems, IT workloads, and grid-level renewable energy. These interactions determine facility-level efficiency metrics such as power usage effectiveness (PUE), water usage effectiveness (WUE), carbon usage effectiveness (CUE), and energy use efficiency (EUE)~\cite{diazmarin2025flipping}. EnergyPlus, a dominant building energy modeling software, is well suited for data center facility-level analyses. EnergyPlus supports both auditing whole building simulation for existing data centers and thereby the planning of future ones. It contains a mature library of plant-side equipment models, including curve-based chillers, cooling towers, dry coolers, pumps, and plant-loop controls. As our work demonstrates, these EnergyPlus components can be reused to serve a liquid-cooled IT load directly rather than being re-implemented. However, EnergyPlus lacks the liquid-cooled IT load object required to interface with these equipment models.

In this work, we develop a scalable liquid cooling object for EnergyPlus. We validate the object against an independent Modelica implementation to assess its accuracy in modeling liquid cooling configurations. Section~\ref{sec:background} reviews existing approaches to liquid cooling modeling and liquid cooling architectures. Section~\ref{sec:method} describes the key calculations applied in the liquid cooling object. Section~\ref{sec:validation} presents the validation against Modelica. Section~\ref{sec:case_study} applies the object in a case study of a mid-sized data center. Section~\ref{sec:discussion} and Section~\ref{sec:conclusion} discuss the limitations and summarize key takeaways.

\begin{table*}[t]
\centering
\small
\renewcommand{\arraystretch}{1.25}
\caption{Data-center thermal simulation tools by modeling scale; this work targets the building/facility scale.}
\label{tab:tool_tiers}
\begin{tabularx}{\textwidth}{@{} p{0.13\textwidth} p{0.12\textwidth} X X @{}}
\toprule
\textbf{Scale} & \textbf{Time scale} & \textbf{Representative tools} & \textbf{Role in the DC liquid-cooling workflow} \\
\midrule
\textbf{Component} & \textmu s--s
  & ANSYS Fluent/Icepak~\cite{ansys_icepak}, Flotherm~\cite{siemens_flotherm}, Cadence Reality DC~\cite{cadence_reality_dc}, OpenFOAM~\cite{openfoam}
  & Design verification of rack layout and heat-transfer surfaces; single-timestep snapshots. Cannot resolve annual operation or building coupling. \\
\textbf{Loop} & s--days
  & Modelica (Buildings Library)~\cite{WetterBuildingsLibrary}, Amesim~\cite{SiemensAmesim}, Simscape Fluids~\cite{mathworks_simscape_fluids}
  & Dynamic response of pumps, valves, CDUs, controllers; start-up transients and control tuning. Facility-scale annual horizons are impractical. \\
\textbf{Building/\allowbreak facility} & 1\,min--years
  & EnergyPlus~\cite{crawley2001energyplus}, TRNSYS~\cite{trnsys2026}, Modelica (Buildings Library)~\cite{WetterBuildingsLibrary}, Spawn-of-EnergyPlus~\cite{Spawn03032024}
  & Integrated thermal, electrical, water, and carbon flows in one engine. The only scale at which facility-boundary efficiency metrics are physically defined and where compliance and regulatory frameworks (e.g., ASHRAE~90.4~\cite{ashrae9042022}, LEED~\cite{usgbc_leed_rating_system}, California Title~24~\cite{cec_2025_title24}) operate. \\
\textbf{Grid} & 5\,min--decades
  & OpenDSS~\cite{epri_opendss}, GridLAB-D~\cite{gridlabd_pnnl}, PLEXOS~\cite{energyexemplar_plexos}, HOMER Pro~\cite{homer_pro}
  & External context: tariffs, emissions factors, demand response, portfolio dispatch. Facility thermal behavior abstracted as a constant or scheduled load. \\
\bottomrule
\end{tabularx}
\end{table*}

\section{Background}
\label{sec:background}

\subsection{EnergyPlus for Data Centers}
EnergyPlus is an open-source building energy simulation engine developed primarily by the U.S. Department of Energy (DOE), currently maintained by the National Laboratory of the Rockies, that models thermal dynamics, heating, ventilation, and air conditioning (HVAC), and energy flows across various building systems. Operating on a nodal framework under the assumption of well-mixed zone air, it traditionally simulates multi-zone energy performance at hourly or sub-hourly timesteps. While the DOE provides a suite of Commercial Prototype Building Models representing major commercial building types~\cite{usdoe2013commercial}, data centers were long absent from these standardized reference models. 

To fill this critical gap, Sun et al. \cite{SUN2021LBNLAIRCOOLED} introduced prototype data center energy models to enable holistic, scalable evaluations of data center cooling efficiency within a standardized building energy modeling environment. Specifically, they modified the \texttt{ITE} object in EnergyPlus, to feature a dedicated flow control mechanism to support the thermal dynamics of air-cooled data center spaces. This air-cooled IT equipment (ITE) object incorporates supply and return approach temperature differences to account for supply air bypass, hot exhaust recirculation, and air containment management strategies. By adjusting the effective supply air volume and return temperature based on these approach temperature differences, EnergyPlus captures realistic air distribution and temperature offsets without violating the underlying nodal heat balance framework. This integration facilitates accurate energy modeling and comparative evaluations across diverse air-cooled data center topologies, cooling strategies, and efficiency standards. 

The impact of that object, however, extends beyond the simulation capability it added, as it established the air-cooled data center as an openly accessible modeling foundation for subsequent research. Because it brings data-center thermal behavior into EnergyPlus, the object can be used with established building-energy toolchains (e.g., OpenStudio, DesignBuilder, Honeybee) and co-simulation interfaces, placing data-center research within whole-building and district-scale energy analysis~\cite{doe_energyplus_overview}. A growing body of work builds on it, including the physics-based and data-driven modeling of data-center energy, water, and thermal dynamics across alternative cooling configurations~\cite{AILI2024122672, CHEN2025124637}, the development of predictive and reinforcement learning based control strategies~\cite{CHEN2026126975,MAHBOD2022119392,10089168,9796122}, and the construction of open benchmark testbeds for sustainable data-center operation~\cite{naug2025sustaindcbenchmarkingsustainabledata}.

Building upon these air-cooled modeling foundations, in this work, we develop a liquid-cooled ITE object in EnergyPlus to extend its capabilities toward high-density, liquid-cooled data center architectures.

\subsection{Existing tools}
\label{sec:existing_tools}
Existing simulation environments that model data-center liquid cooling can be organized along two axes, spatial scale (spanning from chip to loop, facility, and grid) and temporal horizon (microseconds to decades), yielding the four scales summarized in Table~\ref{tab:tool_tiers}. This work contributes primarily at the building/facility scale by enabling long-term, including annual and multi-year, whole-building simulation of DTC liquid cooling in EnergyPlus. Each tier is served by a mature tool family, but no single tier alone can express the whole-building, long-term planning under various scenarios that motivates this work.

At the \textit{component level}, computational fluid dynamics (CFD) tools such as ANSYS Fluent/Icepak~\cite{ansys_icepak}, Siemens FloTHERM~\cite{siemens_flotherm}, Cadence Reality DC~\cite{cadence_reality_dc}, and OpenFOAM~\cite{openfoam} resolve chip, cold-plate, and rack thermofluidics at sub-second timesteps. They are the standard for cold-plate and rack-layout design verification, but their computational cost restricts them to short transient windows and precludes annual or multi-year facility studies.

At the \textit{loop level}, one-dimensional thermo-fluid network tools, including Simcenter Amesim~\cite{SiemensAmesim}, Simscape Fluids~\cite{mathworks_simscape_fluids}, and, most prominently for buildings research, the Modelica ecosystem via the Lawrence Berkeley National Lab (LBNL) Buildings library~\cite{WetterBuildingsLibrary}, resolve the dynamic response of pumps, valves, heat exchangers, and controllers. This tier is well suited for plant-level control sequence tuning and equipment cycling, but scaling it to facility-level analysis is costly, owing to stiff hybrid differential-algebraic equation (DAE) integration, state events from equipment switching, and the translation overhead of detailed envelope models~\cite{WetterBuildingsLibrary, Spawn03032024}. As a result, annual runs of a fully coupled Modelica envelope-plus-plant model are practical for targeted design and control studies, whereas load-based workflows remain faster and better supported for regulatory analysis~\cite{Spawn03032024}.

At the \textit{building level}, the tier occupied by whole-building energy simulators including EnergyPlus~\cite{crawley2001energyplus} and TRNSYS~\cite{trnsys2026}, thermal, electrical, water, and carbon flows are solved in a single engine at one-minute to multi-year horizons. This is the only tier at which the efficiency metrics that govern data-center design and reporting are physically defined, since each is an integral over the facility boundary rather than a loop- or component-local quantity. It is also the tier referenced by building codes and rating systems such as ASHRAE Standard~90.4~\cite{ashrae9042022}, LEED~\cite{usgbc_leed_rating_system}, and California Title~24~\cite{cec_2025_title24}. Modelica-based platforms nominally reach this tier through the Buildings library~\cite{WetterBuildingsLibrary} and the Spawn-of-EnergyPlus coupling~\cite{Spawn03032024}, but their application at the whole-facility scale remains largely research-grade rather than standard engineering practice due to the computational and regulatory constraints noted above.

At the \textit{grid level}, tools such as OpenDSS~\cite{epri_opendss}, GridLAB-D~\cite{gridlabd_pnnl}, PLEXOS~\cite{energyexemplar_plexos}, and HOMER Pro~\cite{homer_pro} provide the external context of tariffs, emissions factors, and demand-response signals, but they abstract each facility as a scheduled or constant load and cannot resolve internal cooling behavior.

Two consequences follow. First, the tools that best resolve liquid-cooling physics are confined to design-tier horizons, which is why liquid-cooled equipment is well characterized at the component and loop levels but essentially absent from operational, regulatory, and carbon-accounting workflows for whole facilities. As recently highlighted by Ganesh et al.~\cite{ganesh_nexus_2026}, this temporal and spatial confinement forces the industry to rely on siloed evaluations that isolate single metrics like energy efficiency (e.g., PUE) while neglecting critical trade-offs within the broader energy-water-carbon nexus and dynamic thermal compliance over annual weather cycles. Second, among building-tier simulators, EnergyPlus is the tool most tightly coupled to plant, envelope, zone air, weather, on-site generation, and grid-emission inputs in a single open-source solver, and it is the reference engine for the code frameworks that govern data-center compliance. Adding a native liquid-cooling object to EnergyPlus therefore fills the specific tier-level gap identified in Table~\ref{tab:tool_tiers}, rather than duplicating capability that already exists elsewhere.

\subsection{Liquid cooling architectures}
\label{sec:liquid_cooling}
Data-center liquid cooling is commonly classified by how close the coolant is brought to the heat source~\cite{AZARIFAR2024123112}. In \emph{liquid-to-room} systems, chilled water feeds CRAH units or in-row coolers that condition room air before it reaches unmodified servers. In \emph{liquid-to-rack} systems, rear-door heat exchangers absorb server exhaust heat at the rack boundary. In \emph{liquid-to-chip} systems, more commonly called \emph{direct-to-chip} (DTC), cold plates are bonded directly to central processing units (CPUs) and graphics processing units (GPUs), with residual air cooling handling lower-power components. A related family, immersion cooling, instead submerges entire servers in dielectric fluid that is either pumped away as a liquid (single-phase) or boiled at the chip surface and condensed overhead (two-phase). Bringing the coolant closer to the die progressively raises the allowable coolant temperature and heat-flux density, often at the cost of greater server modification.

DTC has emerged as the industry standard for the high-density racks that AI workloads require, and immersion, although attractive in theoretical efficiency, is constrained by cost and the difficulty of retrofitting existing air-cooled halls~\cite{WU2025121525}. The operating temperature envelope of any liquid-cooled plant is codified by the 2021 ASHRAE thermal guidelines for liquid cooling~\cite{ashrae2024reference}, which partition facility water supply temperatures into six classes (Table~\ref{tab:liquidclass}). Higher classes permit warmer water and unlock free cooling and waste-heat recovery, but require chip packages and cold plates rated for the elevated temperature.

To demonstrate the capability and performance of the developed EnergyPlus object across DTC cooling applications, we select the two extremes of the ASHRAE envelope as representative operating points: \textbf{W17} for conventional chilled-water designs, and \textbf{W45} for warm-water designs that leverage free-cooling and heat-reuse potential. These two classes bracket the design space in which the proposed model is validated and applied.

\begin{table}
\centering
\caption{Facility water supply temperatures by liquid cooling class, from the
2021 ASHRAE thermal guidelines for liquid cooling~\cite{ashrae2024reference}. The
minimum water temperature for all classes is 2\textdegree C (35.6\textdegree F).}
\label{tab:liquidclass}
\begin{tabular}{@{} lcc @{}}
\toprule
\multirow{2}{*}{\textbf{Liquid Cooling Class}} & \multicolumn{2}{c}{\textbf{Facility Water Supply Temperature}} \\
\cmidrule(lr){2-3}
 & \textbf{\textdegree C} & \textbf{\textdegree F} \\
\midrule
W17 & 17    & 62.6   \\
W27 & 27    & 80.6   \\
W32 & 32    & 89.6   \\
W40 & 40    & 104    \\
W45 & 45    & 113    \\
W+  & $>$45 & $>$113 \\
\bottomrule
\end{tabular}
\end{table}

\section{Method}
\label{sec:method}
\begin{figure}[h]
	\centering
	\includegraphics[width=.95\columnwidth]{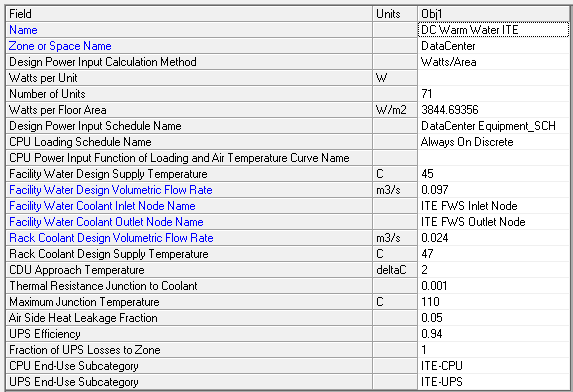}
	\caption{Screenshot of the developed ElectricEquipment:ITE:LiquidCooled object in EnergyPlus in the IDF editor.}
	\label{fig:ep_object}
\end{figure}

To model DTC liquid cooling of IT equipment within
a whole-building energy simulation, we develop a custom EnergyPlus
object, \texttt{ElectricEquipment:\allowbreak ITE:\allowbreak LiquidCooled}, as a native
plant-loop component. Throughout our modeling process we focus on single-phase DTC cooling due to its operational maturity and widespread commercial adoption~\cite{WU2025121525}. As summarized in Table~\ref{tab:fields} and shown
in Figure~\ref{fig:ep_object}, the object is derived from the existing
\texttt{ITE}
object ~\cite{SUN2021LBNLAIRCOOLED} in three steps. 

First, we retain the fields that describe the IT load itself, such as the server zone, IT power density, load schedules, CPU power curve, and uninterruptible power supply (UPS) efficiency and loss fraction. For our case of liquid cooling, the amount of heat that goes to the zone air compared to the liquid loop is controlled by the air leakage fraction parameter. The CPU power curve is a biquadratic function of the loading fraction $x$ (0--1, from the loading schedule) and the rack coolant supply temperature $y = T_\mathrm{rack,sup}$ (\textdegree C). The existing EnergyPlus air-cooled object~\cite{SUN2021LBNLAIRCOOLED, energyplus2026inputoutput} uses the same curve with the server inlet air temperature as $y$, and the curve coefficients are user inputs. \citet{SUN2021LBNLAIRCOOLED} obtained such coefficients by fitting server measurements from \citet{Sahini2017} and found that CPU power is nearly independent of inlet temperature, i.e., approximately proportional to load. Although the field names retain ``CPU'' from the air-cooled object, the modeled load is the rack's electric power, so the object applies equally to CPU, GPU, or other accelerator racks. All simulations in this work leave the curve unspecified, in which case the object defaults to CPU power linear in load.

Second, we remove the air-side fields that have no
meaning for a liquid loop, such as the server-fan power and airflow
curves, the design entering-air temperature, and the supply-air node.
Third, we add ten new fields that define the liquid-cooling path. The
shared IT-load fields are reused with no changes to obtain the instantaneous
CPU electric power $P_\mathrm{cpu}$ (W) and the UPS heat loss
$Q_\mathrm{ups,loss}$ (W), where $Q_\mathrm{ups,loss}$ denotes the
portion of UPS input power dissipated as heat inside the served zone.
\begin{table}
  \small
  \caption{Fields for the air- and liquid-cooled objects in EnergyPlus 26.1. A checkmark marks applicability.}
  \label{tab:fields}
  \begin{tabularx}{\columnwidth}{X c c}
    \toprule
    \textbf{Field} & \textbf{Air} & \textbf{Liq.} \\
    \midrule
    \multicolumn{3}{l}{\emph{Fan / airflow}}\\
    Fan power fraction              & \checkmark & \\
    Fan airflow curves             & \checkmark & \\
    Fan power curve                & \checkmark & \\
    Design entering air temperature& \checkmark & \\
    Supply air node                & \checkmark & \\
    Approach temperature schedules & \checkmark & \\
    Fan end-use subcategory        & \checkmark & \\
    \addlinespace
    \multicolumn{3}{l}{\emph{Server \& IT load}}\\
    Server zone                    & \checkmark & \checkmark \\
    IT power density               & \checkmark & \checkmark \\
    IT load schedule               & \checkmark & \checkmark \\
    CPU loading schedule           & \checkmark & \checkmark \\
    CPU power curve                & \checkmark & \checkmark \\
    CPU end-use subcategory        & \checkmark & \checkmark \\
    \addlinespace
    \multicolumn{3}{l}{\emph{UPS}}\\
    UPS end-use subcategory        & \checkmark & \checkmark \\
    UPS efficiency                 & \checkmark & \checkmark \\
    Fraction of UPS losses to zone & \checkmark & \checkmark \\
    \addlinespace
    \multicolumn{3}{l}{\emph{Facility / rack coolant loop}}\\
    Facility Water System (FWS) supply temp. & & \checkmark \\
    FWS flow rate                  & & \checkmark \\
    FWS inlet node                 & & \checkmark \\
    FWS outlet node                & & \checkmark \\
    Rack coolant flow rate         & & \checkmark \\
    Rack coolant supply temperature& & \checkmark \\
    CDU approach temperature difference       & & \checkmark \\
    \addlinespace
    \multicolumn{3}{l}{\emph{Chip thermal}}\\
    Junction $R_{\text{th}}$ (chip thermal resistance) & & \checkmark \\
    Maximum junction temperature   & & \checkmark \\
    Air leakage fraction to room air & & \checkmark \\
    \bottomrule
  \end{tabularx}
\end{table}

\begin{figure*}[t]
	\centering
	\includegraphics[width=\textwidth]{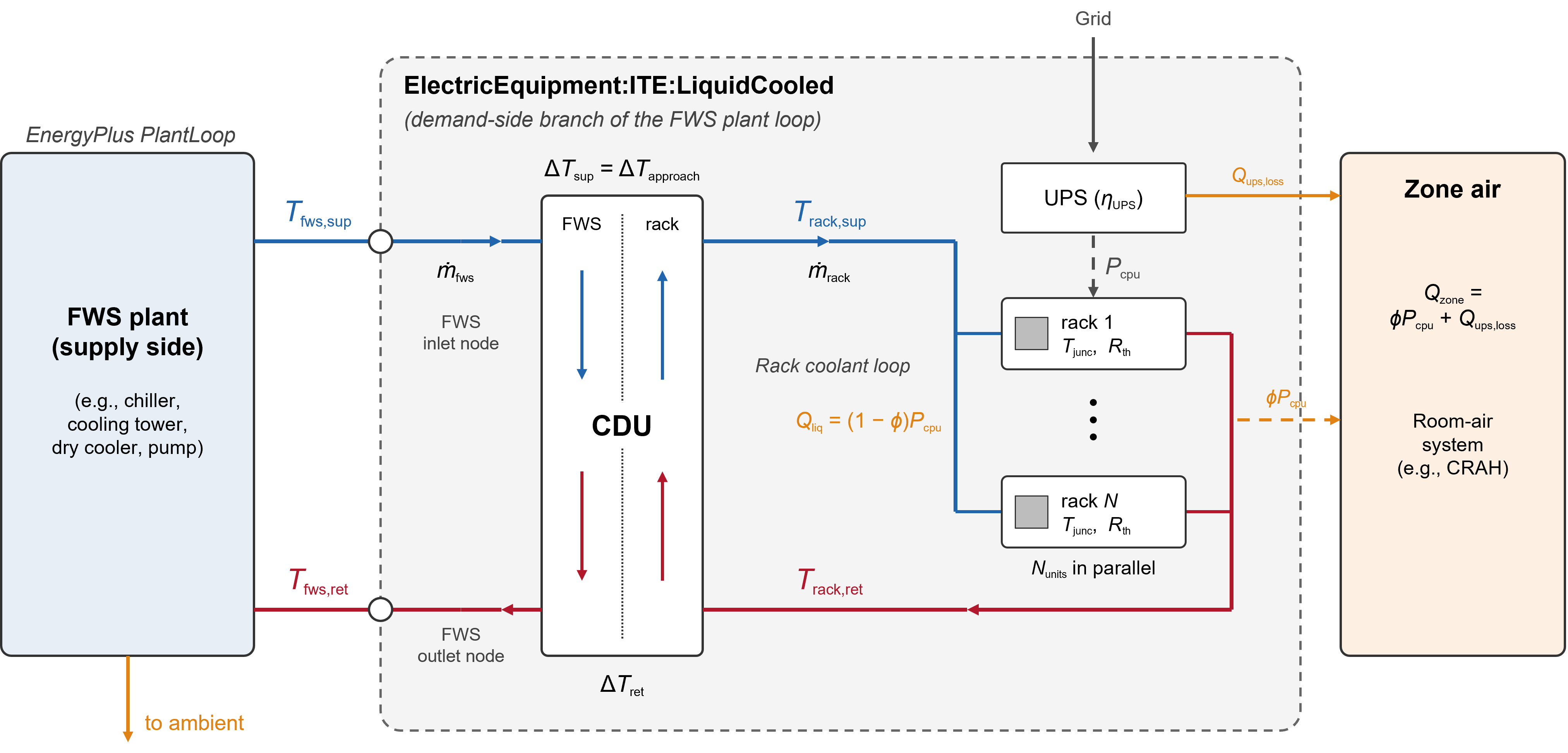}
	\caption{Schematic of the \texttt{ElectricEquipment:\allowbreak ITE:\allowbreak LiquidCooled} object and its coupling to the FWS plant loop and the server zone.}
	\label{fig:schema}
\end{figure*}

Whereas the air-cooled object rejects all IT heat to the zone air, the
liquid-cooled object registers on the demand side of an EnergyPlus
\texttt{PlantLoop}, the Facility Water System (FWS), so that the heat
captured by the cold plates is delivered directly to the building's chilled- or warm-water plant and handled by the plant's supply-side equipment (e.g., chiller, cooling tower, or dry cooler). The added cooling path is deliberately quasi-steady and algebraic: at each zone timestep the object evaluates a closed-form chain of energy balances rather than iterating a discretized coolant network, keeping the object numerically robust and cheap at annual sub-hourly resolution while still resolving the temperatures that govern DTC feasibility. It is decomposed into three sequentially solved stages: first, a heat-partition step that splits the processor heat between coolant and room air; second, a primary FWS loop modeled as a real plant branch; and third, an algebraic rack-side secondary loop and CDU. Since the object is defined as a demand-side component on the FWS loop, facility-level heat rejection is automatically handled by EnergyPlus's existing plant infrastructure. Figure~\ref{fig:schema} shows the resulting loop topology and the temperatures defined in the following subsections.

\subsection{Heat partition between coolant and room air}

The first new field, the air-leakage fraction $\phi \in [0,1]$
(dimensionless), controls how the processor (e.g., CPU, GPU, TPU) heat, which for the air-cooled object goes entirely to the zone, is split into a liquid-captured heat rate
$Q_\mathrm{liq}$ (W) and an air-leaked contribution to the zone gain
$Q_\mathrm{zone}$ (W):
\begin{equation}
    Q_\mathrm{liq} = (1-\phi)\,P_\mathrm{cpu}, \qquad
    Q_\mathrm{zone} = \phi\,P_\mathrm{cpu} + Q_\mathrm{ups,loss},
    \label{eq:partition}
\end{equation}
where $P_\mathrm{cpu}$ (W) is the instantaneous CPU electric power obtained from the retained IT-load fields and $Q_\mathrm{ups,loss}$ (W) is the inherited UPS heat loss defined above. $Q_\mathrm{liq}$ is passed to the plant loop, while $Q_\mathrm{zone}$, the leaked CPU heat plus the inherited UPS losses, is registered as a convective internal gain on the zone-air heat balance, where it is picked up by whatever room-air system serves the space. The air-leakage fraction is the single lumped parameter through which the
model represents all non-ideal thermal capture (bypass, radiation,
uncaptured server-fan heat) of a real cold-plate rack.

\subsection{Primary facility-water loop}

Four new fields define the primary loop: the FWS design supply temperature $T_\mathrm{fws,sup}^{\mathrm{des}}$ (\textdegree C), the FWS design volumetric flow rate $\dot V_\mathrm{fws}$
(m\textsuperscript{3}\,s\textsuperscript{-1}), and the FWS coolant inlet and outlet node names, which connect the object as a demand-side branch of the plant loop. The inlet/outlet nodes make the FWS side the only part of the model that is a true fluid network. At sizing, the design mass flow rate $\dot m_\mathrm{fws}$ (kg\,s\textsuperscript{-1}) is derived from the flow-rate field and the temperature-dependent coolant density $\rho_\mathrm{fws}$ (kg\,m\textsuperscript{-3}) evaluated at the design supply temperature,
\begin{equation}
    \dot m_\mathrm{fws}
    = \rho_\mathrm{fws}\!\left(T_\mathrm{fws,sup}^{\mathrm{des}}\right)\,\dot V_\mathrm{fws},
    \label{eq:mfws}
\end{equation}
and this flow is requested from the plant each timestep. The design supply-temperature field sets the initial condition at the start of each simulation run, but the operating FWS supply temperature $T_\mathrm{fws,sup}$ (\textdegree C) is then read from the inlet node, that is, set by the upstream plant equipment and controls, not prescribed
internally. The FWS return temperature $T_\mathrm{fws,ret}$ (\textdegree C) follows from a first-law energy balance on the branch, 
\begin{equation}
    T_\mathrm{fws,ret} = T_\mathrm{fws,sup}
        + \frac{Q_\mathrm{liq}}{\dot m_\mathrm{fws}\,c_{p,\mathrm{fws}}},
    \label{eq:tfwsret}
\end{equation}
where $c_{p,\mathrm{fws}}$ (J\,kg\textsuperscript{-1}\,K\textsuperscript{-1})
is the coolant specific heat capacity, re-evaluated at the current supply temperature. A guard handles the zero-flow case: if the loop delivers no flow, no heat can be extracted and $Q_\mathrm{liq}$ is redirected to the zone-air gain, preventing a non-physical temperature rise. The outlet node temperature and mass flow are written back for the plant solver to close the loop, so the
liquid load is rejected consistently with the rest of the building model and its part-load plant controls.

\subsection{CDU and rack secondary loop}

Three new fields define the CDU and secondary loop: the rack coolant design supply temperature $T_\mathrm{rack,sup}^{\mathrm{des}}$ (\textdegree C), the
rack coolant design volumetric flow rate $\dot V_\mathrm{rack}$
(m\textsuperscript{3}\,s\textsuperscript{-1}), and the CDU approach
temperature difference $\Delta T_\mathrm{approach}$ (K), a user input. The CDU, that is, the liquid-to-liquid heat exchanger separating facility water from rack coolant, is not modeled as a rated heat exchanger. Instead the user-specified approach temperature difference $\Delta T_\mathrm{approach}$ (K) is applied directly to give the operating rack coolant supply temperature ($T_\mathrm{rack,sup}$ (\textdegree C)), following the approach-temperature method introduced for the air side of EnergyPlus data center models~\cite{SUN2021LBNLAIRCOOLED}:
\begin{equation}
    T_\mathrm{rack,sup} = T_\mathrm{fws,sup} + \Delta T_\mathrm{approach}.
    \label{eq:tracksup}
\end{equation}
This is a key simplification of the liquid path: it pins the
rack-coolant supply a constant offset above the facility-water supply, independent of load, and therefore does not capture the growth of approach temperature as the CDU duty rises toward its UA limit. The rack coolant return temperature $T_\mathrm{rack,ret}$ (\textdegree C) is closed by a first-law balance on the constant-design-flow secondary loop,
\begin{gather}
    \dot m_\mathrm{rack} = \rho_\mathrm{rack}\!\left(T_\mathrm{rack,sup}^{\mathrm{des}}\right)\,\dot V_\mathrm{rack}, \\
    T_\mathrm{rack,ret} = T_\mathrm{rack,sup} + \frac{Q_\mathrm{liq}}{\dot m_\mathrm{rack}\,c_{p,\mathrm{rack}}},
    \label{eq:trackret}
\end{gather}
where $\dot m_\mathrm{rack}$ (kg\,s\textsuperscript{-1}) is the rack
coolant mass flow rate, $\rho_\mathrm{rack}$ (kg\,m\textsuperscript{-3}) its density, and $c_{p,\mathrm{rack}}$
(J\,kg\textsuperscript{-1}\,K\textsuperscript{-1}) its specific heat. The rack coolant design supply-temperature field sets the temperature at which the rack coolant properties are evaluated, which keeps the rack mass flow rate at its design value, and the initial rack coolant temperatures at the start of each simulation period. For modeling simplicity, the loop is assumed to use the same working fluid as the FWS. For reporting only, the object also computes the counter-flow log-mean temperature difference across the CDU $\Delta T_\mathrm{lm}$ (K), 

\begin{gather}
    \Delta T_\mathrm{lm} = \frac{\Delta T_1 - \Delta T_2}{\ln(\Delta T_1/\Delta T_2)}, \\
    \Delta T_1 = T_\mathrm{rack,ret}-T_\mathrm{fws,ret}, \quad
    \Delta T_2 = T_\mathrm{rack,sup}-T_\mathrm{fws,sup},
    \label{eq:lmtd}
\end{gather}

with $\Delta T_1$ and $\Delta T_2$ (K) the local temperature differences at the two ends of the CDU. The expression reduces to the arithmetic mean in the balanced-flow limit. It does not feed back into the temperature solution. We report the log-mean temperature difference to verify if the heat exchanger assumed is realistic.

\subsection{Chip-junction temperature}

The final two new fields, the junction-to-coolant thermal resistance
$R_{\mathrm{th}}$ (K\,W\textsuperscript{-1}) and the maximum allowable
junction temperature $T_\mathrm{junc}^{\max}$ (\textdegree C), govern DTC
thermal feasibility. The junction-to-coolant thermal resistance combines chip packaging, thermal interface materials, and conduction path with liquid convection. It is referenced to the coolant entering the rack and absorbs liquid temperature rise along the cold plate, so its value depends on flow rate. The peak chip-junction temperature $T_\mathrm{junc}$
(\textdegree C) is estimated with a single lumped thermal-resistance
relation,
\begin{equation}
    T_\mathrm{junc} = T_\mathrm{rack,sup}
        + \frac{Q_\mathrm{liq}}{N_\mathrm{units}}\,R_{\mathrm{th}},
    \label{eq:tjunc}
\end{equation}
where $N_\mathrm{units}$ is the number of rack units on the secondary loop. The rack units are assumed to be connected in parallel, each receiving coolant at $T_\mathrm{rack,sup}$ and an equal share of $Q_\mathrm{liq}$. Flow within a rack is not resolved, and any series arrangement of cold plates is lumped into $R_\mathrm{th}$. The loop temperatures are computed from the total flow and are independent of this assumption. Our calculation treats each rack as one equivalent thermal element between coolant and junction. It does not resolve cold-plate convection or die-level spreading, and $R_{\mathrm{th}}$ is a fixed, load-independent constant. The object accumulates the hours for which $T_\mathrm{junc}$ exceeds $T_\mathrm{junc}^{\max}$, giving a compliance metric without altering the energy solution.

\subsection{Efficiency metrics calculation}
To evaluate the overall sustainability and thermal energy recovery performance of the proposed data center framework, we utilize a combination of key efficiency metrics. To fully capture both the traditional infrastructure overhead and the added value of waste heat utilization, we calculate the PUE, WUE, and EUE, as defined in Equations~\eqref{eq:pue} through~\eqref{eq:eue}~\cite{LBNL2026,diazmarin2025flipping}. While PUE is widely used as the standard to quantify the ratio of total facility energy to IT equipment energy, it inherently treats all rejected heat as a loss and does not adequately reflect the benefits of thermal recovery. Similarly, WUE evaluates site water consumption per unit of IT energy, providing critical insight into cooling-related resource demands and potential trade-offs between energy and water efficiency~\cite{LBNL2024}. To address the limitation of PUE, we incorporate EUE, a newly proposed metric designed to incentivize waste heat utilization. By accounting for both the energy directed into compute components and the reused thermal energy, EUE provides a more comprehensive assessment, potentially allowing the calculated efficiency to exceed 1.0~\cite{diazmarin2025flipping}.

\begin{align}
  \mathrm{PUE} &= \frac{E_{\mathrm{facility}}}{E_{\mathrm{IT}}}, \qquad
  E_{\mathrm{facility}} = E_{\mathrm{IT}} + E_{\mathrm{UPS}} + E_{\mathrm{cool}} + E_{\mathrm{light}},
  \label{eq:pue} \\[4pt]
  \mathrm{WUE} &= \frac{V_{\mathrm{w}}}{E_{\mathrm{IT}}}, \label{eq:wue} \\[4pt]
  \mathrm{EUE} &= \frac{E_{\mathrm{compute}} + E_{\mathrm{reused}}}{E_{\mathrm{facility}}}. \label{eq:eue}
\end{align}

In these definitions, $E_{\mathrm{IT}}$ is the IT electricity at the UPS
output (including server fans in the air-cooled case), $E_{\mathrm{UPS}}$ the
UPS conversion loss ($\eta_{\mathrm{UPS}} = 0.94$), $E_{\mathrm{cool}}$ the
electricity of chillers, heat-rejection equipment, pumps, and air-handling fans, and
$E_{\mathrm{light}}$ interior lighting. Distribution losses downstream of the
UPS are not modeled. $V_{\mathrm{w}}$ (L) is the total on-site water consumption, including cooling tower evaporation, drift, and blowdown as well as any adiabatic cooling or humidification water, so that WUE is expressed in L\,kWh\textsuperscript{-1} with $E_{\mathrm{IT}}$ in kWh.  $E_{\mathrm{compute}}$ is the processor electricity and $E_{\mathrm{reused}}$ the recovered liquid-loop heat.

\begin{figure}[t] 
	\centering
	\includegraphics[width=\columnwidth]{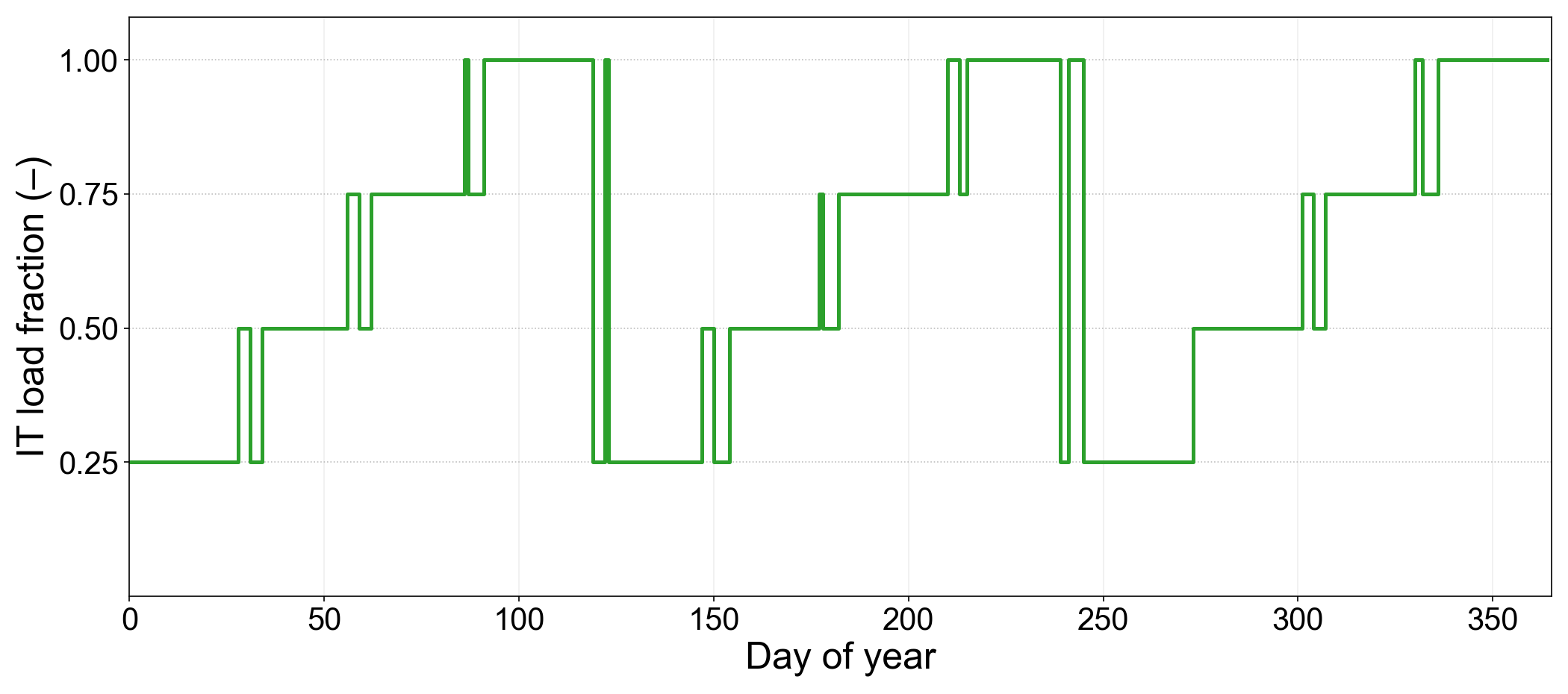} 
	\caption{Annual IT-load schedule: a daily step schedule with four discrete utilization levels (0.25, 0.50, 0.75, 1.00) arranged as repeating quarterly ramps. The CPU power scales linearly with the load fraction, reaching the 2,143~kW design value at full load.}
	\label{fig:it_load}
\end{figure}

\begin{table*}[t]
\centering
\caption{Model parameters for the two plant configurations. The upper block is common to the validation and case-study models; the lower block lists what changes between them. Spanning values are shared by both plants; split entries differ.}
\label{tab:validation_params}
\begin{tabular}{@{} l cc @{}}
\toprule
\textbf{Parameter} & \textbf{Chilled Water (W17)} & \textbf{Warm Water (W45)} \\
\midrule
\multicolumn{3}{@{}l}{\textit{Common to the validation and case-study models}} \\
\addlinespace
Data-hall floor area            & \multicolumn{2}{c}{557.42\,m\textsuperscript{2}} \\
IT power density                & \multicolumn{2}{c}{3\,844.69\,W\,m\textsuperscript{-2}} \\
Peak CPU electrical power       & \multicolumn{2}{c}{2\,143.10\,kW} \\
Number of rack units            & \multicolumn{2}{c}{71} \\
UPS efficiency                  & \multicolumn{2}{c}{0.94} \\
Facility-water volumetric flow  & \multicolumn{2}{c}{0.097\,m\textsuperscript{3}\,s\textsuperscript{-1}} \\
CDU approach temperature difference        & \multicolumn{2}{c}{2.0\,K} \\
Facility-water supply setpoint  & 12.0\,\textdegree C & 45.0\,\textdegree C \\
Rack-coolant supply temperature & 14.0\,\textdegree C & 47.0\,\textdegree C \\
Heat rejection (liquid loop)    & 2.5\,MW chiller (rated reference COP 5.5) + wet tower & 3.0\,MW dry cooler \\
\addlinespace
\midrule
\multicolumn{3}{@{}l}{\textit{Changed from the validation setup to the case study}} \\
\addlinespace
Air-side heat-leakage fraction  & \multicolumn{2}{c}{0.05 (validation) $\rightarrow$ 0.20 (case study)} \\
Zone-air heat handling          & \multicolumn{2}{c}{Ideal-loads system (validation) $\rightarrow$ CRAH four-pipe fan-coil (case study)} \\
CRAH cooling source\textsuperscript{a} & Main chilled-water loop & Dedicated air-cooled chiller $+$ room CHW loop \\
\bottomrule
\end{tabular}

\vspace{2pt}
{\footnotesize \textit{\textsuperscript{a}Case-study only; the validation
models carry no CRAH and dissipate the 5\,\% leakage through the ideal-loads
system.}}
\end{table*}

\section{Validation}\label{sec:validation}
\subsection{Setup}

To verify the correctness of the newly developed EnergyPlus object
\texttt{ElectricEquipment:ITE:LiquidCooled}, which models a DTC-cooled IT load
together with its CDU and facility-water interface, we cross-validate it against
the widely adopted
Modelica \emph{Buildings} library~\cite{WetterBuildingsLibrary} (LBNL Buildings
13.0.0, solved in OpenModelica 1.26.3). We developed the Modelica model
independently as an equation-based, component-level reconstruction of the same
plant, reproducing the governing relations (IT heat balance, CDU approach,
and loop energy balances) in a mature simulator
solved by a different numerical method. Agreement between the two
implementations therefore supports the correctness of the EnergyPlus object
rather than merely its internal self-consistency.

The comparison spans 21 operating conditions, covering the parameter ranges
over which liquid-cooled data centers operate. The facility-water supply
temperature is varied from
12 to 45\,\textdegree C, reaching the defining maximum of every ASHRAE
liquid-cooling class up to W45 (Table~\ref{tab:liquidclass}); the rack-coolant
flow is varied to give rack temperature rises from
10.3 to 41.0\,K; the CDU approach from
4 to 10\,K; the facility-water flow from half to twice its design value
(0.097\,m\textsuperscript{3}\,s\textsuperscript{-1}); and the IT load from zero
to the full 2,143\,kW design power. Two of the conditions are realized as
complete plants and simulated over a full year: a chilled-water plant in which
an electric chiller backed by a wet cooling tower supplies 12\,\textdegree C water, and a warm-water plant in
which a dry cooler rejects the liquid load directly to ambient air at
45\,\textdegree C (Table~\ref{tab:validation_params}). In the remaining 19
conditions the facility-water boundary condition is imposed directly, which is
what allows each design parameter to be varied independently and over a range
no single heat-rejection plant could deliver.

All conditions share the same data hall and the same IT side, simulated using
the Typical Meteorological Year 3 (TMY3) weather dataset for Tampa, FL, USA,
provided by EnergyPlus. The two annual runs are driven by the IT-utilization
schedule of the DOE prototype data center model~\cite{SUN2021LBNLAIRCOOLED}
(Figure~\ref{fig:it_load}); detailed parameters are summarized in
Table~\ref{tab:validation_params}. The air-side heat-leakage fraction is set to
0.05 in both models so that the liquid heat load is identical. The 5\,\% air
portion itself lies outside the validation scope because the Modelica model
represents only the liquid side, while EnergyPlus removes this heat from the
zone with an idealized air system
(\texttt{ZoneHVAC:\allowbreak IdealLoadsAirSystem}) that supplies exactly the cooling required without modeling any fans, coils, or chillers. Neither model
therefore contains an air-side equipment model to compare. Matching the
fraction nonetheless ensures the comparison isolates the liquid cooling plant
rather than confounding it with a differing heat-split assumption.

The two full-plant conditions were simulated over a full year in both tools.
EnergyPlus ran at a one-minute timestep and reported every plant variable at
that resolution, whereas the Modelica replica was integrated as a single
continuous annual run with a variable-step solver (relative tolerance
$10^{-6}$, cold-initialized once at the start of the year) and written out on
the same one-minute grid. The validation is therefore carried out at a
one-minute timestep over the full year. The Modelica models use
\texttt{Buildings.Media.\allowbreak Specialized.\allowbreak Water.\allowbreak
TemperatureDependentDensity} as the water medium rather than the library's default, 
constant-property \texttt{Buildings.\allowbreak Media.\allowbreak Water}, so that density follows
temperature as it does in EnergyPlus.

The Modelica warm-water plant is cold-initialized once at the start of the
January-1 run, whereas EnergyPlus begins from its converged annual state. Its
facility-water controller therefore needs about the first hour of the year to
settle from this cold start, producing a start-up transient that is absent in
EnergyPlus. We exclude the first hour of the year from the comparison for this
reason, leaving 525,541 paired samples per variable. In the chilled-water
plant the supply temperature is set directly by the chiller rather than through
a controller loop, so it has no such start-up transient. The first hour is
nevertheless dropped from both plants, so that the two comparisons rest on an
identical set of samples.

Sweeping the facility-water setpoint inside a complete plant is not viable: the
dry cooler cannot hold a 32\,\textdegree C supply against Tampa summer air, and
driving the electric chiller to such a leaving-water temperature evaluates its
performance curves far outside the range over which they were fitted, so an
equipment limitation would surface as a residual of the object under test.

For the remaining 19 conditions the heat-rejection equipment is therefore
replaced by an ideal source in both tools, a
\texttt{PlantComponent:\allowbreak TemperatureSource} on the EnergyPlus supply branch and a
prescribed-temperature, prescribed-flow source in Modelica, leaving the pump,
the plant loop, the demand branch and the object itself unchanged. In each
condition the IT load is stepped from zero to full in eight levels, and
statistics are taken over the last hour of each level, once the loop has
settled (480 samples per condition). One design parameter is varied at a time.

For each variable we report the root-mean-square error (RMSE), the mean
absolute error (MAE), and the maximum absolute error (Max) between the two
simulators, with the sign convention bias $=$ Modelica $-$ EnergyPlus.
Reporting the MAE and the maximum alongside the RMSE separates a genuinely
large residual from one caused by a few isolated timesteps. Aligned with the
precision guidelines~\cite{hodder2024measurement}, a 0.1\,K
threshold is used to validate the simulated temperatures.

\subsection{Full-year agreement}
\begin{figure}[t]
  \centering
  \includegraphics[width=\linewidth]{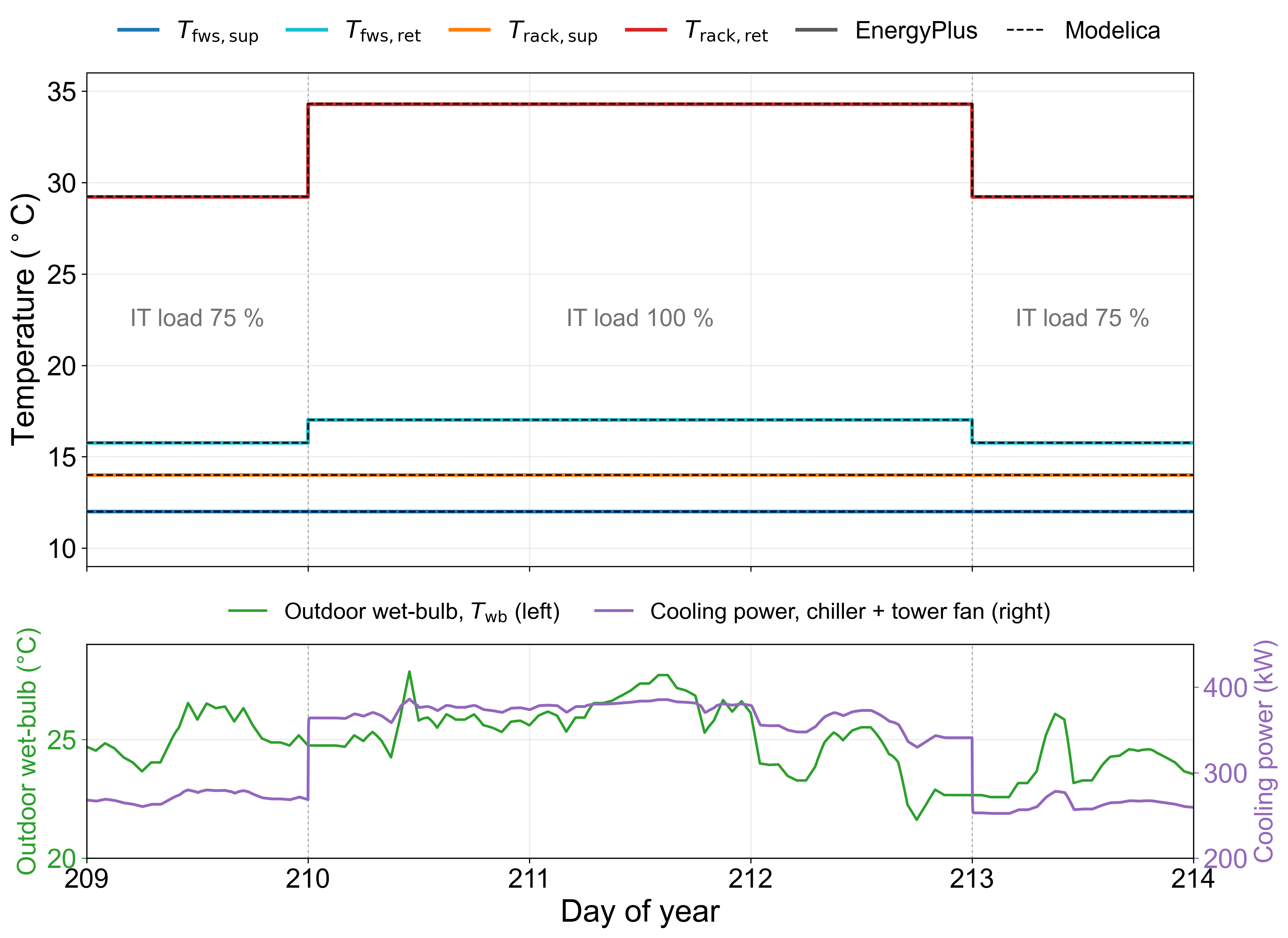}\\[2pt]
  \includegraphics[width=\linewidth]{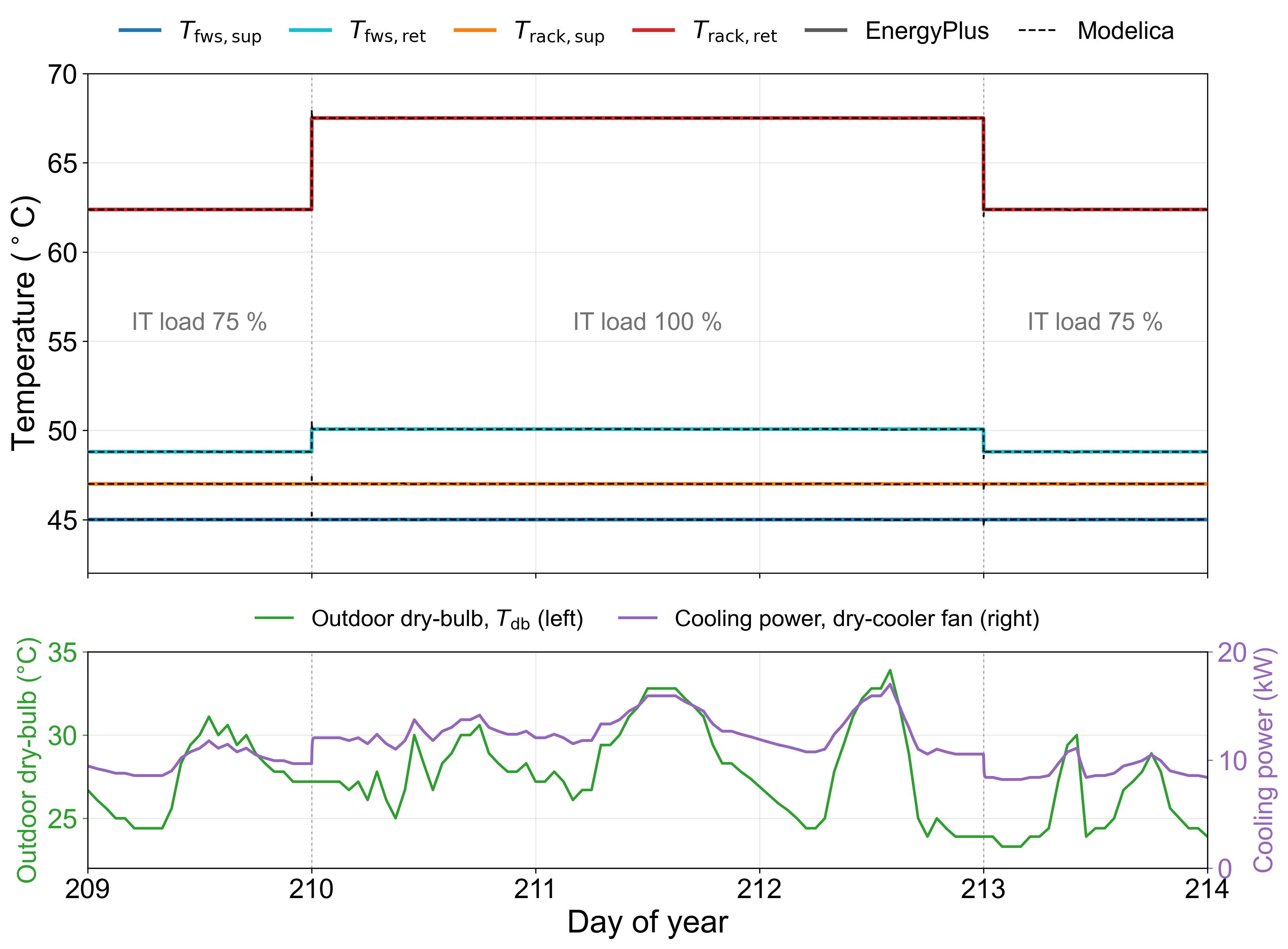}
  \caption{EnergyPlus and Modelica loop temperatures for the chilled-water (top) and warm-water (bottom) plants over days 209--214, with IT utilization labeled. The two traces are indistinguishable. Lower panels: EnergyPlus outdoor air temperature and cooling power, shown for context and not part of the comparison.}
  \label{fig:validation_timeseries}
\end{figure}

The full-year agreement between the EnergyPlus object and the Modelica baseline
is summarized in Table~\ref{tab:validation_independent}, computed at one-minute
intervals with the first hour of the year excluded. Additionally,
Figure~\ref{fig:validation_timeseries} illustrates a detailed snapshot of the
dynamic thermal response over a representative multi-day period for both the
chilled-water and warm-water configurations. The steps on days 210 and 213
follow the IT-utilization schedule (75\,\% to 100\,\% and back). Only the
return temperatures ($T_\mathrm{fws,ret}$, $T_\mathrm{rack,ret}$) change. The
supply temperatures ($T_\mathrm{fws,sup}$, $T_\mathrm{rack,sup}$) are fixed by
the plant setpoint and the 2\,K CDU approach temperature difference, while the
return temperatures rise with the extracted heat,
$Q_\mathrm{liq}/(\dot m\,c_p)$. The rack-coolant temperature difference
($T_\mathrm{rack,ret}-T_\mathrm{rack,sup}$) increases from 15.2 to 20.3\,K,
about four times the facility-water increase
($T_\mathrm{fws,ret}-T_\mathrm{fws,sup}$, 3.8 to 5.0\,K), because the rack flow
is about one quarter of the facility-water flow. The cooling power in the lower
panels, in contrast, follows the outdoor air temperature ($T_\mathrm{db}$ for
the warm-water plant, $T_\mathrm{wb}$ for the chilled-water plant) as well as
the IT load (8.6 to 17.0\,kW for the dry-cooler fan and 260 to 386\,kW for the
chiller and cooling tower), because the plant absorbs the weather to hold the
supply temperature at its setpoint.

\begin{table*}[t]
\centering
\caption{Full-year agreement between the EnergyPlus liquid-cooled object and
the Modelica baseline for the warm-water and chilled-water plants (1-min
intervals, first hour excluded, 525,541 samples per variable). Variables that
agree to numerical precision in both plants are omitted.}
\label{tab:validation_independent}
\begin{tabular*}{\textwidth}{@{\extracolsep{\fill}} l ccc ccc @{}}
\toprule
\multirow{2}{*}{\textbf{Variable}} & \multicolumn{3}{c}{\textbf{Chilled water (W17)}} & \multicolumn{3}{c}{\textbf{Warm water (W45)}} \\
\cmidrule(lr){2-4} \cmidrule(lr){5-7}
 & \textbf{RMSE} & \textbf{MAE} & \textbf{Max} & \textbf{RMSE} & \textbf{MAE} & \textbf{Max} \\
\midrule
Facility-water supply temperature            & 0.000 & 0.000 & 0.000 & 0.008 & 0.003 & 1.362 \\
Facility-water return temperature            & 0.004 & 0.003 & 0.005 & 0.008 & 0.004 & 1.363 \\
Rack-coolant supply temperature              & 0.000 & 0.000 & 0.000 & 0.008 & 0.003 & 1.362 \\
Rack-coolant return temperature              & 0.021 & 0.019 & 0.030 & 0.008 & 0.003 & 1.362 \\
Facility-water mass flow rate (kg\,s\textsuperscript{-1}) & 0.002 & 0.002 & 0.002 & 0.001 & 0.001 & 0.001 \\
CDU log-mean temperature difference          & --    & --    & --    & 0.001 & 0.001 & 0.001 \\
\bottomrule
\end{tabular*}
\end{table*}

Across the annual simulation, all four loop temperatures agree closely in both
plants, with RMSE at or below 0.021\,K and MAE at or below
0.019\,K. In the chilled-water plant, the facility-water and
rack-coolant supply temperatures coincide to 0.000\,K (RMSE, MAE and
Max), because both simulators hold the 12\,\textdegree C setpoint directly
rather than through a controller loop. The facility-water return temperature
agrees to an RMSE of 0.004\,K (Max 0.005\,K), and the
rack-coolant return temperature is the largest chilled-water residual at
0.021\,K RMSE (MAE 0.019\,K, Max 0.030\,K).
The warm-water plant agrees to 0.008\,K RMSE and 0.003\,K MAE on every loop
temperature, supply and return alike, and to 0.001\,K on the CDU log-mean
temperature difference.

The rack-coolant return temperature carries the largest sustained residual,
0.021\,K RMSE in the chilled-water plant. In the warm-water plant it is no
larger than the supply-side temperatures. Because the rack supply temperature
matches to 0.000\,K (both models pin it at the facility-water supply
temperature plus the 2\,K approach), this residual enters entirely through the
return-side heat balance, where the rack return temperature is the rack supply
temperature plus the injected heat divided by the rack-coolant heat-capacity
rate. The Modelica model uses a constant water specific heat of
4,180\,J\,kg\textsuperscript{-1}\,K\textsuperscript{-1}, whereas EnergyPlus
takes it from its temperature-dependent water-property tables. That constant is
the value for water near 47\,\textdegree C, so the difference vanishes at the
warm-water design point and grows toward colder supply temperatures, which is
what the two plants show. The mismatch therefore scales with the injected heat,
and hence with IT load, and is independent of ambient temperature, identifying
it as a fluid-property evaluation difference rather than a controller artifact.
It remains well within tolerance.

\begin{figure}[t]
  \centering
  \includegraphics[width=\linewidth]{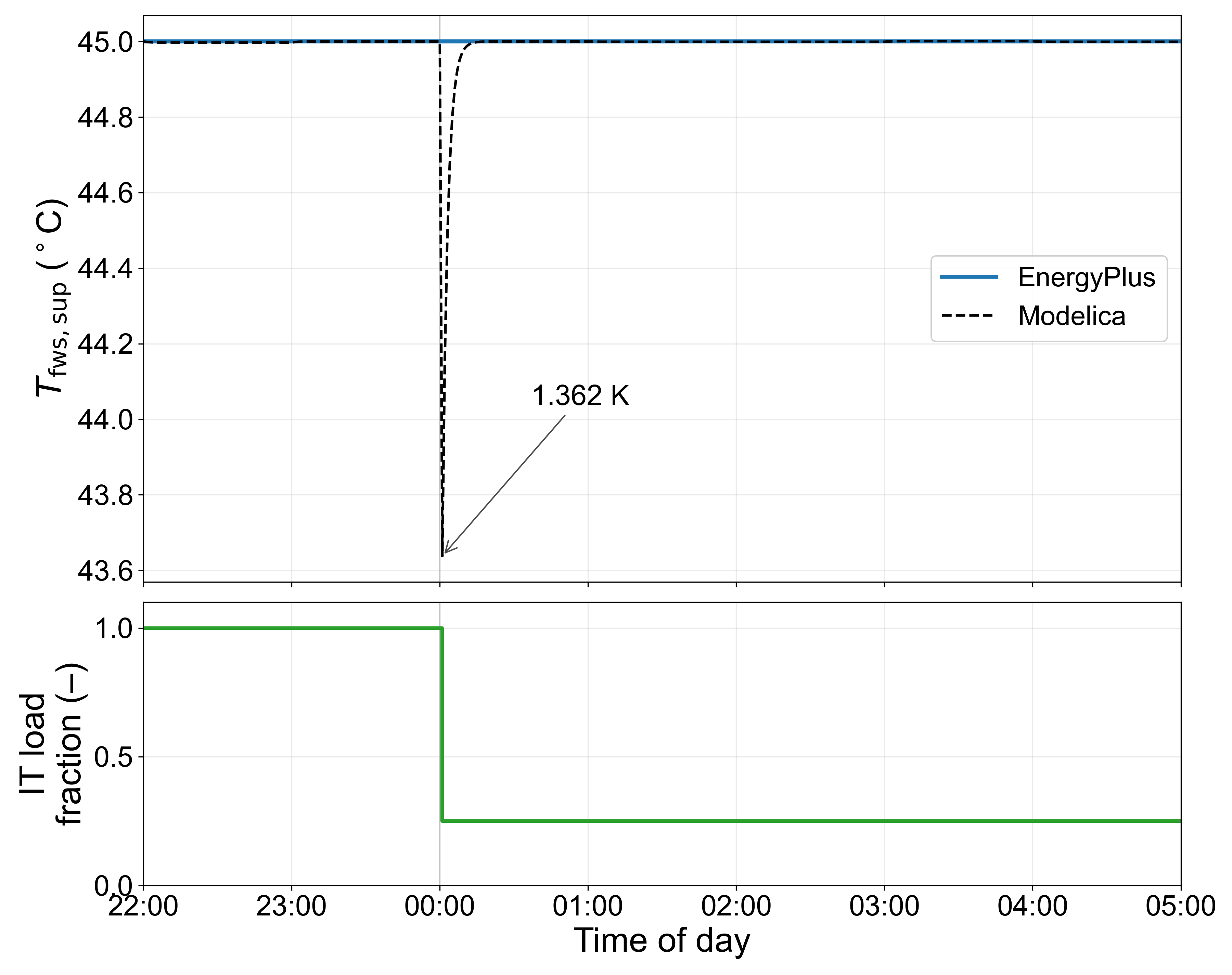}
  \caption{Detailed view of the temperature residual spike during an isolated
  load step, illustrating the transient discrepancy between EnergyPlus and
  Modelica due to control-modeling differences.}
  \label{fig:loadstep}
\end{figure}

The only residual exceeding this tolerance is the warm-water maximum, which
reaches 1.362\,K. Because the average error is far smaller than the maximum
(MAE 0.003\,K, RMSE 0.008\,K), this residual comes from a few isolated
timesteps rather than a sustained offset, and its origin is a difference in
control-modeling scope rather than plant physics, as illustrated in
Figure~\ref{fig:loadstep}.

EnergyPlus reads the facility-water supply temperature from the plant node,
which its solver drives to the 45\,\textdegree C setpoint quasi-statically
within each timestep, so the delivered temperature tracks the setpoint without
controller lag. The Modelica plant instead resolves the dry-cooler fan
proportional--integral (PI) loop explicitly (integral time 120\,s). At each
midnight IT-load step the heat rejected to the loop changes abruptly, and the
finite-rate controller briefly overcools the water before the supply
temperature recovers exponentially, returning to within 0.1\,K of the setpoint
in under 7\,min. These excursions occur only at the load steps: every
one-minute sample whose error exceeds 0.1\,K falls on one of the 31 days on
which the utilization schedule changes level, and the chilled-water plant,
whose setpoint is enforced directly by the chiller, shows no such transient at
all. The deviation is therefore not a plant-physics disagreement but a
consequence of the two tools' modeling scope. EnergyPlus idealizes the plant as
tracking its setpoint instantaneously each timestep, whereas the Modelica model
captures the finite-rate response of the physical fan controller.

\subsection{Sensitivity across operating conditions}
\label{sec:validation_conditions}

Table~\ref{tab:validation_conditions} reports the remaining 19 conditions. The
facility-water and rack-coolant supply temperatures agree to 0.000\,K in every
condition. The residual is confined to the two return temperatures and the CDU
log-mean temperature difference. 

\begin{table*}[t]
\centering
\caption{Agreement across the 19 swept conditions, reported for the same
variables and in the same order as Table~\ref{tab:validation_independent}. Each
entry is the worst value over the conditions of that group (settled samples, 480
per condition; K unless noted). Variables that agree to numerical precision in
every condition are omitted, as in Table~\ref{tab:validation_independent}.}
\label{tab:validation_conditions}
\begin{tabular*}{\textwidth}{@{\extracolsep{\fill}} l cc cc cc cc @{}}
\toprule
\multirow{3}{*}{\textbf{Variable}}
 & \multicolumn{2}{c}{\textbf{FWS temperature}}
 & \multicolumn{2}{c}{\textbf{Rack-coolant flow}}
 & \multicolumn{2}{c}{\textbf{CDU approach}}
 & \multicolumn{2}{c}{\textbf{FWS flow}} \\
 & \multicolumn{2}{c}{12 to 45\,\textdegree C}
 & \multicolumn{2}{c}{$\Delta T_\mathrm{rack}$ 10 to 41\,K}
 & \multicolumn{2}{c}{4 to 10\,K}
 & \multicolumn{2}{c}{0.5--2$\times$ design} \\
 & \multicolumn{2}{c}{9 conditions} & \multicolumn{2}{c}{4 conditions}
 & \multicolumn{2}{c}{4 conditions} & \multicolumn{2}{c}{2 conditions} \\
\cmidrule(lr){2-3} \cmidrule(lr){4-5} \cmidrule(lr){6-7} \cmidrule(lr){8-9}
 & \textbf{RMSE} & \textbf{Max} & \textbf{RMSE} & \textbf{Max}
 & \textbf{RMSE} & \textbf{Max} & \textbf{RMSE} & \textbf{Max} \\
\midrule
Facility-water supply temperature & 0.000 & 0.000 & 0.000 & 0.000 & 0.000 & 0.000 & 0.000 & 0.000 \\
Facility-water return temperature & 0.004 & 0.006 & 0.003 & 0.005 & 0.003 & 0.005 & 0.007 & 0.011 \\
Rack-coolant supply temperature   & 0.000 & 0.000 & 0.000 & 0.000 & 0.000 & 0.000 & 0.000 & 0.000 \\
Rack-coolant return temperature   & 0.019 & 0.030 & 0.001 & 0.002 & 0.008 & 0.014 & 0.001 & 0.001 \\
Facility-water mass flow rate (kg\,s\textsuperscript{-1}) & 0.014 & 0.014 & 0.011 & 0.011 & 0.011 & 0.011 & 0.023 & 0.023 \\
CDU log-mean temperature difference & 0.004 & 0.007 & 0.001 & 0.002 & 0.005 & 0.007 & 0.002 & 0.003 \\
\bottomrule
\end{tabular*}
\end{table*}

The largest residual over the whole sweep is 0.030\,K, on the rack-coolant
return temperature at the coldest facility-water supply temperature
(0.019\,K RMSE at 12\,\textdegree C), consistent with the constant specific
heat identified above. The facility-water return temperature stays at or below
0.011\,K and the CDU log-mean temperature difference at or below 0.007\,K,
while the facility-water mass flow rate agrees to
0.023\,kg\,s\textsuperscript{-1} or better. Every entry is within the
0.1\,K tolerance.

\subsection{Validity and scope}

Over 21 operating conditions, covering every ASHRAE liquid-cooling class from
W17 to W45, rack-coolant temperature rises of 10 to 41\,K, CDU
approaches of 4 to 10\,K, facility-water flows from half to
twice design, and IT loads from zero to full, every sustained temperature
residual is at or below 0.030\,K. Each of these residuals has an identified
cause in the Modelica reference model, either the constant specific heat it
uses or the dry-cooler fan controller it resolves explicitly where EnergyPlus
idealizes the plant. Neither is an error in the EnergyPlus object under test.

The validation does not cover the zone-air model or the routing of air-side heat leakage, which in the production models is larger and is handled by a CRAH outside the liquid-side comparison. It also does not cover the heat-rejection plant and its electricity for chillers, pumps, and
fans, or full annual energy integrals. The CDU is modeled at a fixed, user-specified approach rather than with a load-dependent effectiveness, the secondary rack loop is represented algebraically at constant mass flow, and all runs use a single climate.

\begin{figure*}[t] 
	\centering
	\includegraphics[width=1.0\textwidth]{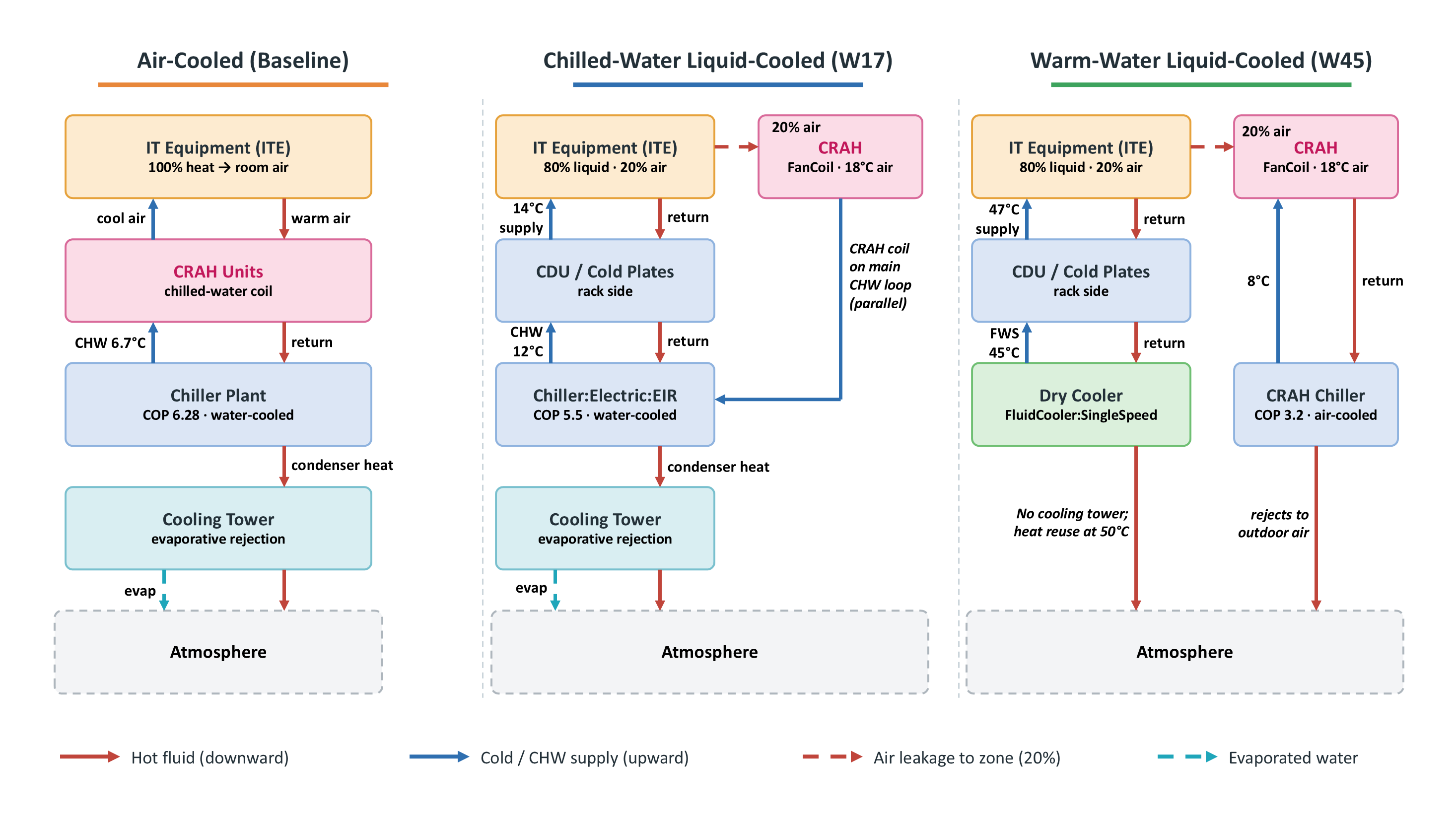}
	\caption{Schematic comparison of the three cooling architectures modeled in EnergyPlus for the case study. The air-cooled baseline relies on standard object types, whereas both the chilled-water and warm-water systems utilize our newly developed liquid-cooling object.}
	\label{fig:configs}
\end{figure*}

\section{Case Study: Tampa, FL, USA}
\label{sec:case_study}
To demonstrate the capabilities of the developed EnergyPlus liquid-cooling object and its evaluation through key efficiency metrics (PUE, WUE, and EUE), we conduct the case study for a data center located in Tampa, Florida, USA, a hot, humid climate case, driven by the same TMY3 weather dataset used for validation over a full annual run. To evaluate the energy performance, we compare a total of three cooling configurations: an air-cooled baseline and two DTC liquid-cooled architectures (chilled-water and warm-water), as illustrated in Figure~\ref{fig:configs}. 

The case-study models are the validated models of Section~\ref{sec:validation} with one addition. The validation models set the air-side heat-leakage fraction to 0.05 and remove that heat with an ideal-loads system, whereas the case study raises the fraction to 0.20 and adds a CRAH to remove the resulting room load. In practice, DTC liquid cooling may capture greater than 70\% of rack heat load~\cite{vertiv2026transforming}, while room air removes the remainder, so a 0.20 air-side fraction reflects this split more realistically. The CRAH is a variable-flow four-pipe fan-coil unit supplying air at 18\,\textdegree C. All other parameters remain those of Table~\ref{tab:validation_params}. All three configurations, including the air-cooled baseline, use the same UPS efficiency (0.94) and lighting power density (6.89\,W/m$^2$).

Because the object releases the air-side fraction into the thermal zone as an ordinary internal gain, the CRAH and its chilled-water supply are added entirely within the input file, reusing EnergyPlus's existing zone heat balance, fan-coil, and plant-loop components without further development. Implementing the liquid-cooled object natively in EnergyPlus therefore places hybrid air- and liquid-cooled data centers within the whole-building workflows (envelope, weather, schedules, sizing, and code-compliance modeling) in which the program is already established.

The two liquid-cooled architectures serve the CRAH coil differently. In the warm-water model the 45\,\textdegree C facility water is too warm to make cold supply air, so a small air-cooled chiller feeds a separate room chilled-water loop that serves the coil. In the chilled-water model the CRAH coil is placed directly on the main chilled-water loop, because the main chiller already delivers cold water, so no extra chiller is needed.

\begin{figure*}[t] 
	\centering
	\includegraphics[width=.9\textwidth]{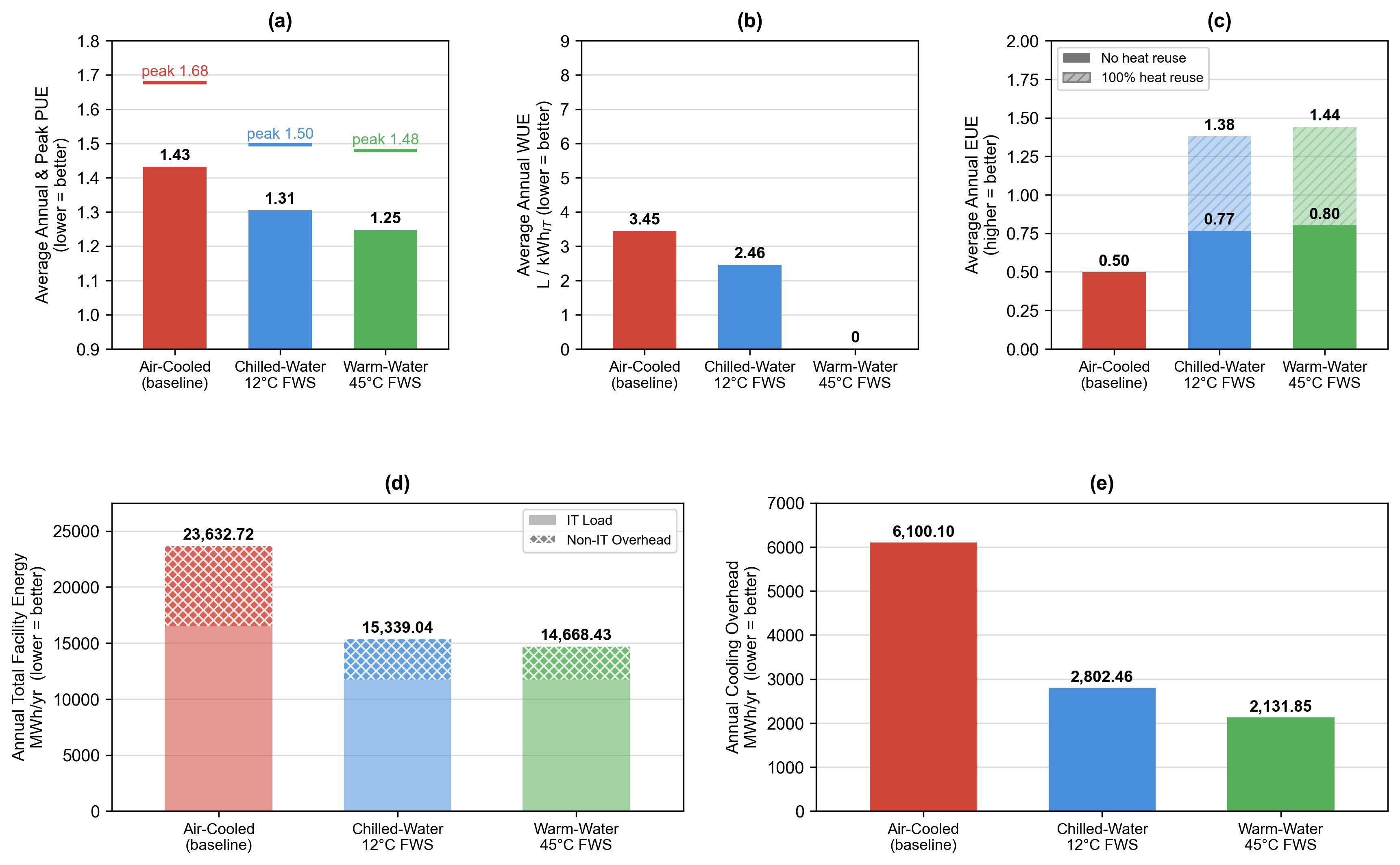}
	\caption{Annual PUE, WUE, EUE, facility energy, and cooling overhead of the air-cooled baseline and the chilled-water (12\,\textdegree C) and warm-water (45\,\textdegree C) liquid-cooled configurations in Tampa, FL.}
	\label{fig:metrics}
\end{figure*}

As shown in Figure~\ref{fig:metrics}, the chilled-water and warm-water liquid cooling systems show significant advantages over the air-cooled baseline for all key metrics calculated. To summarize the trends between the air-cooled baseline and warm water loop, the annual PUE drops from 1.43 to 1.25 as the WUE goes from 3.45 to 0 and EUE increases from 0.50 to 0.80, or to 1.44 if the liquid-loop heat is fully reused (Figure \ref{fig:metrics} a-c). For reference, the air-cooled baseline PUE is close to the 2023 U.S. average PUE of approximately 1.4~\cite{LBNL2024}, and both liquid-cooled cases fall within the range of approximately 1.07 to 1.33 simulated for a direct liquid-cooled facility~\cite{GHENI2026PUE}. Reductions in site-level water use follow from avoiding evaporative heat rejection~\cite{LEI2025WUE}. The full-reuse EUE value assumes 100\% reuse of the liquid-loop heat extraction (equivalent to approximately 80\% of the IT input) as recoverable waste heat. In practice the usable fraction depends on the temperature at which the heat is delivered and on the requirements of the downstream application, so these values represent a theoretical upper bound for applications such as district heating rather than an expected yield. Additionally, we observe that the peak PUE drops from 1.68 to 1.48 from the air cooled baseline to the warm water loop for liquid cooling. The warm water loop achieves this sharp reduction by eliminating the chiller plant and cooling tower from the liquid loop, allowing for efficient heat dissipation through dry cooling to occur. Lastly, we observe that the total facility energy reduces by 37.9\% (23,633 to 14,668 MWh/yr) and cooling overhead drops by 65.1\% (6,100 to 2,132 MWh/yr), confirming the high energy efficiency gains from DTC liquid cooling. Of the 8,964\,MWh/yr total reduction, 53.1\,\% comes from the IT side itself, as DTC cooling removes the server-fan power that the air-cooled baseline requires, 44.3\,\% from the cooling plant, and the remaining 2.7\,\% from the correspondingly smaller UPS losses.

Interestingly, the warm-water system does not outperform the chilled-water system across all hours. The warm-water hourly PUE is higher than that of the chilled-water system for 785 hours of the year (9.0\%), exclusively at the minimum 25\% IT load and in cool, dry weather (mean wet-bulb 12.3~\textdegree C compared with 19.0~\textdegree C across all 25\%-load hours), where the dry cooler's benefit scales down with the liquid heat while the chilled-water plant's water-cooled chiller is at its most efficient. Furthermore, the air-cooled chiller serving the room load in the warm-water plant operates at a lower COP than the water-cooled chiller in the chilled-water plant, and this trade-off would not occur if both configurations shared the same chiller efficiency. The gap in those hours never exceeds 0.023 in PUE. Both directions of this trade-off are climate-driven. Tampa's hot, humid summer limits the free-cooling potential of the dry cooler relative to colder regions, while its mild winter is what produces the crossover hours at all, so the margin between the two liquid-cooled configurations should be expected to shift with location. Evaluating both configurations across a range of ASHRAE climate zones is left to future work. 

\section{Discussion and Limitations}
\label{sec:discussion}

The primary contribution of this work is a validated object for whole-building simulation of DTC liquid cooling inside EnergyPlus. The custom \texttt{ElectricEquipment:\allowbreak ITE:\allowbreak LiquidCooled} object embeds the distinctive physics of a two-loop DTC system inside the same solver that already handles the building envelope, zone-air balance, and central plant. It represents the facility-water primary loop, the rack-side secondary loop, the CDU that couples them, the liquid/air split of the chip heat, and the junction-to-coolant thermal resistance. Liquid-cooled data centers can therefore be simulated natively in EnergyPlus at the same whole-building level of resolution previously available only for air-cooled data centers. This makes it possible to study a broader range of cooling-system configurations with high fidelity in applications such as data-center siting and grid coordination, alongside conventional air-cooled designs within a common modeling framework.

Because the object is now available inside EnergyPlus, the validated plant-level results support a broad class of building- and facility-level studies. These include design-point sizing of DTC cooling plants, annual energy comparisons of alternative plant configurations across climate zones and data-center sizes, and supervisory scheduling of IT load at hourly or coarser timescales where each interval can be treated as quasi-steady.

Beyond these, the object also exposes the plant-side thermal and flow quantities needed to compute a broader set of efficiency metrics beyond the widely used PUE, including WUE and the recently proposed EUE \cite{LBNL2026,diazmarin2025flipping}. Because EUE credits reused thermal energy, this provides a common basis on which emerging technologies such as waste-heat recovery and district-heat coupling can be quantitatively compared against conventional designs. Combined with EnergyPlus's standard plant components and time-varying grid emission factors, the object further supports sustainability analyses and can serve as a physically grounded input to upstream grid, siting, and scheduling tools. Additionally, although our current models use constant facility-water temperature setpoints, EnergyPlus natively supports dynamic temperature and load-based control strategies. Our liquid cooling module can therefore be applied for both data center planning and supervisory control of cooling operations. 

Despite these benefits, the proposed liquid-cooled object has limitations that bound its validated use. First, the CDU and the rack-side hydraulics are represented algebraically at a single design point. The coolant delivery temperature is set from a fixed, user-specified CDU approach temperature difference rather than a load-dependent heat exchanger. The secondary coolant flow rate is held constant, and the chip-to-coolant path is collapsed into a single fixed junction-to-coolant resistance rather than a resolved cold-plate model. Because the approach of a real CDU scales with the heat it transfers, a fixed approach over-predicts it at part load and biases the rack-coolant and junction temperatures upward. The formulation also cannot represent heterogeneous rack populations or variable secondary-pump operation. Second, the object is steady-state and omits thermal mass and hydraulic dynamics, so it does not resolve the transient response of the liquid loop, as the load-step comparison of Section~\ref{sec:validation} illustrates. It is therefore unsuited to transient or real-time control studies, or to pre-cooling strategies that depend on the capacitance of the loop itself. Third, the reported validation covers only the liquid-side coupling between the two models. The zone air balance lies outside the scope of this comparison, and the object's air-side heat leakage path relies on existing EnergyPlus objects. Last but certainly not least, the case study covers a single data center in one location. The validation itself spans a wide range of operating conditions, but every run uses the same weather file, so the energy and efficiency metric results reported here are specific to Tampa's climate. Future studies should extend the case study across climate zones, where the trade-off between warm-water and chilled-water cooling is expected to be most sensitive.

\section{Conclusion}
\label{sec:conclusion}
Our work adds the ability to model liquid-cooled data centers in EnergyPlus through introducing a new object, \texttt{ElectricEquipment:\allowbreak ITE:\allowbreak LiquidCooled}, for both chilled-water and warm-water DTC cooling architectures. We validate the object against an independently developed Modelica model with matching parameters across 21 operating conditions: full-year chilled-water (W17) and warm-water (W45) plants compared at one-minute resolution, and 19 further conditions spanning facility-water supply temperatures from 12 to 45~\textdegree C, rack-coolant flow, CDU approach, and facility-water flow. Loop-temperature RMSE stays at or below 0.021~K throughout, well inside the 0.1~K acceptance threshold, confirming fidelity for whole-building simulation. In a case study for a mid-sized data center in Tampa, FL, USA, the warm-water loop reduced PUE from 1.43 to 1.25, eliminated on-site water use, and cut total facility energy by 37.9\% relative to an air-cooled baseline. By embedding liquid-cooling logic directly in EnergyPlus, this contribution extends whole-building energy modeling practice in both academia and industry as data center infrastructure is rapidly shifting toward DTC liquid cooling systems for AI data centers, all within a single open-source framework.

\bibliographystyle{ACM-Reference-Format}
\section*{Data Availability Statement}
The source code is available at \url{https://github.com/Urban-Informatics-Lab/energyplus-liquid-cooling}.

\section*{Declaration of Interests}
The authors declare that they have no conflict of interest.

\section*{Declaration of Generative AI Usage}
Claude (Sonnet 4.6) was used to help write the data analysis and EnergyPlus object code. Claude (Sonnet 4.6) was also used to help improve the readability and language of this manuscript and to develop Figure 6. The author(s) reviewed and edited all content as needed and take full responsibility for the content of the publication.
\begin{acks}
This work was supported in part by the Precourt Institute for Energy (PIE) and the Center for Integrated Facility Engineering (CIFE). Any opinions, findings, and conclusions or recommendations expressed in this material are those of the authors and do not necessarily reflect the views of PIE and/or CIFE. The authors acknowledge the contributions of Shreya Ramachandran to the preliminary ideation of the project.
\end{acks}

\bibliography{ep-dc-liquid-cooling}

\end{document}